\documentclass[aps,prl,twocolumn,groupedaddress]{revtex4}
\usepackage{epsfig}
\DeclareMathAlphabet{\mathitb}{OT1}{cmr}{bx}{sl}
\begin{document}

\renewcommand{\thefootnote}{\fnsymbol{footnote}}
\title{NMR Probing Spin Excitations in the Ring-Like Structure of
a Two-Subband System}
\author{X. C. Zhang}
\email{xczhang@physics.ucla.edu}
\author{G. Scott}
\author{H. W. Jiang}

\affiliation{Department of Physics and Astronomy, University
of California at Los Angeles, 405 Hilgard Avenue, Los Angeles,
CA 90095, USA}

\date{\today}

\begin{abstract}

Resistively detected nuclear magnetic resonance (NMR) is observed
inside the ring-like structure, with a quantized Hall conductance
of 6$e^2/h$, in the phase diagram of a two subband electron system.
The NMR signal persists up to 400 mK and is absent in other states
with the same quantized Hall conductance.  The nuclear spin-lattice
relaxation time, $T_1$, is found to decrease rapidly towards the
ring center. These observations are consistent with the assertion
of the ring-like region being a ferromagnetic state that is
accompanied by collective spin excitations.

PACS numbers: 73.43.Nq, 71.30.+h, 72.20.My
\end{abstract}

\maketitle

A two-dimensional electron system consisting of two filled subbands is
emerging as an experimental laboratory to study charge and spin
correlation effects.  The correlations become particularly prominent when the
two sets of Landau levels with different subband quantum numbers are
brought into degeneracy by varying magnetic field and/or carrier concentration.
A series of experimental observations were made to explore the consequences of
real electron spin exchange interactions \cite{Zhang2005PRL, Ellen2005PRB},
as well as pseudospin charge excitations in the vicinity of the
degeneracy regions \cite{ Muraki2001PRL, Zhang2006PRB}.

One of the interesting findings in the recent studies of the
two-subband system is that the experimental phase diagram, in
the density-magnetic field plane, exhibits
pronounced "ring-like structures" at even integer filling
factors \cite{Zhang2005PRL}.
It was conjectured that these structures represent
ferromagnetic phases.
This conjecture was based on theoretical predictions that
a ferromagnetic phase transition can occur when two
Landau levels (LLs) with opposite spin are close
to crossing at the Fermi energy. By promoting all the
electrons, or holes, in the uppermost occupied LL to the
lowest unoccupied LL of opposite spin, the total energy of the system
can be lowered by the exchange mechanism \cite{Giuli1985PRB}.
Such a transition has been seen in a single subband two dimensional hole
system when two LL levels with opposite spin and different Landau level
index approach each other by increasing the Zeeman energy using a
tilted magnetic field \cite{Dane1997PRL}. The ring-like structures
have also been observed recently by Ellenberger \it{et al.}$\;$\rm
on a two-subband parabolic quantum well \cite{Ellen2005PRB}.
The authors, however, suggest that the ring structures can be
single particle states with an enhanced exchange interaction
within each subband in the framework of mean-field theory.

To address the question whether the ring structures are collective
states in nature, measurements other than the conventional transport
are needed. The resistively detected NMR technique has recently
emerged as an effective method to probe collective spin states in the
fractional quantum Hall regime \cite{Smet2002Nature, Stern2004PRB},
the Skyrmion spin texture close to the filling
factor $\nu=1$ \cite{Desrat2002PRL, Gervais2005PRL},
and the role of electron spin polarization in the phase
transition of a bilayer system \cite{Spielman2005PRL}.
This method is significantly less demanding than the
high-sensitivity conventional NMR detection \cite{Barrett1995PRL}.
In this paper we have adapted this technique to study NMR in the
vicinity of the ring structure with a quantized Hall
conductance of $6e^2/h$.  It reveals that the NMR signal is
phase-space dependent and only appears inside the ring structure.
An array of observations suggest that the ring-like region is a
collective state that has intriguing spin excitations.

\begin{figure}[t]
\begin{center}
\includegraphics[width= 80mm, angle=0]{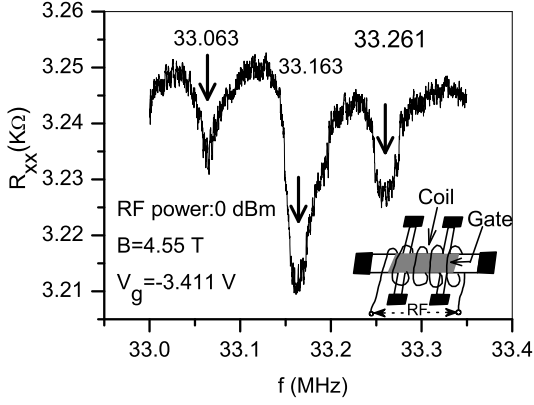}
\end{center}
\vspace{-5mm}
\caption{$^{75}$As NMR signal detected at $B=4.55$ T and $V_g=-3.411$ V,
close to filling factor $\nu=1$ and in a regime of a single
occupied subband.  Resonance frequencies are labeled. The center
resonance is accompanied by two side quadrupole split dips. The inset
shows the gated Hallbar sample around which several turns of
coil were wound for passing RF signal.}
\label{Fig-calibration}
\end{figure}

The sample used in this study is a symmetrical modulation-doped
single quantum well. Both well and spacer are 240 \AA\, thick.
The total electron density at gate
voltage $V_g=0$ is $8.1 \times 10^{11}$ cm$^{-2}$, with a
distribution of 5.4  and $2.7 \times 10^{11}$ cm$^{-2}$ in
the first and second subband, respectively. The mobility is
$4.1 \times 10^5$ cm$^2$/V$\cdot$s.  A NiCr top gated
100 $\mu$m wide Hall bar with 270 $\mu$m between voltage probes was
patterned by standard lithography techniques.
As shown in the inset picture of Fig.~\ref{Fig-calibration},
several turns of NMR coil were wound around the sample, which
was placed in a dilution refrigerator with a base temperature of 60 mK,
and in a perpendicular magnetic field normal to the sample plane.
A small RF magnetic field generated by the coil with a matching
frequency $f=\gamma H_0$ will cause NMR for $^{75}$As nuclei,
where the gyromagnetic ratio $\gamma=7.29$ MHz/T.
The resistance was measured using quasi-dc lock-in techniques
with $I=100$ nA and $f=13.2$ Hz.

To calibrate the external magnetic field resistively
detected NMR close to $\nu=1$ was performed, as shown
in Fig.~\ref{Fig-calibration}. The NMR manifests itself as a dip
in the resistance versus frequency curve due to the increased
electron Zeeman energy caused by depolarizing the nuclei at resonant
conditions \cite{Desrat2002PRL}, see discussions below.
The main dip at 33.163 MHz is about
100 KHz separated with two side ones, which are caused by quadrupole
splitting resultant from the interaction of the nuclear electric
quadrupole moment with an electric field gradient.
Here, the splitting is about three times larger than
in Ref.~\cite{Desrat2002PRL}, possibly due to the large
gate voltage applied across the quantum well.
By comparing the frequency of the center peak with
its theoretical value, the uncertainty of the magnetic field
was determined within $\pm$0.1\%.

\begin{figure}[ht]
\begin{minipage}[t]{0.5\linewidth}
\centering
\epsfig{figure=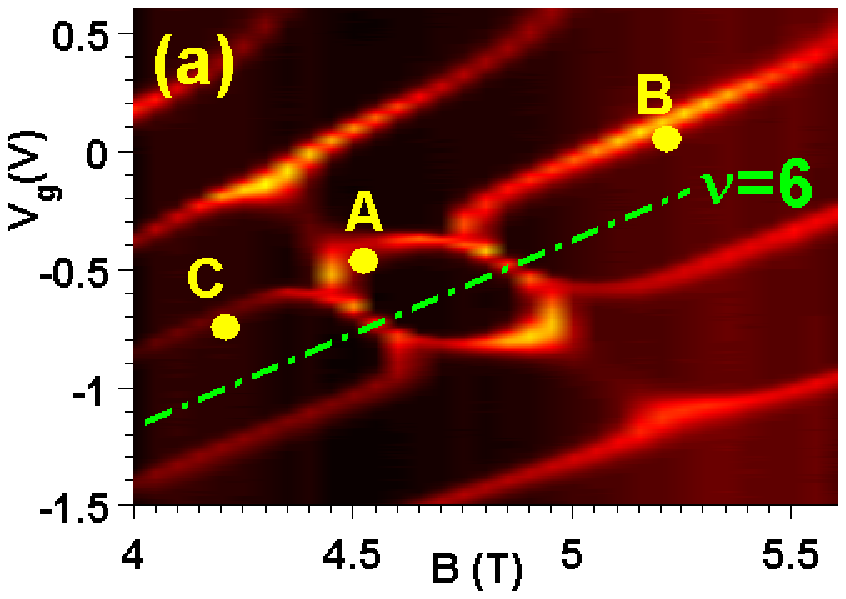, width=1.05\linewidth, bbllx=175,%
bblly=286, bburx=436, bbury=474}
\label{Fig-ring}
\end{minipage}\hfill
\begin{minipage}[t]{0.5\linewidth}
\centering
\epsfig{figure=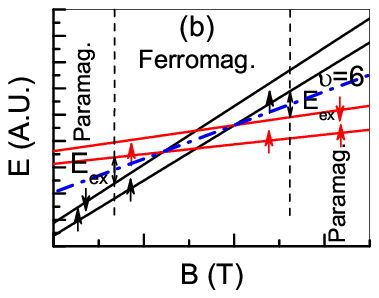, width=1.02\linewidth}
\label{Fig-ring}
\end{minipage}
%\qquad
\centering\epsfig{figure=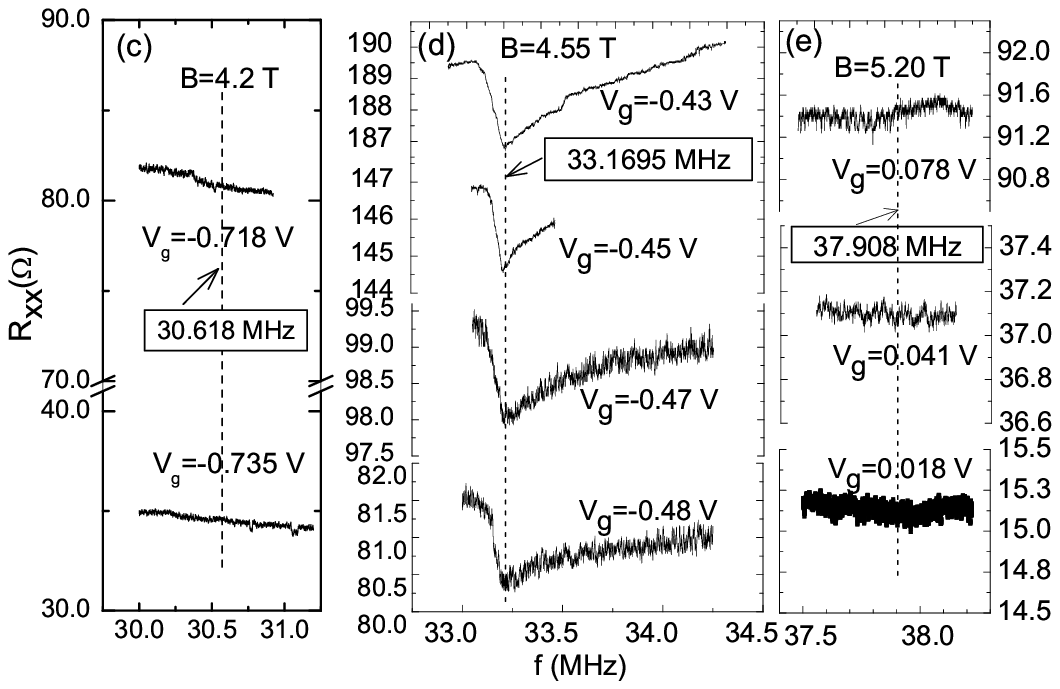, width=1.1\linewidth,%
bbllx=25, bblly=-30, bburx=338, bbury=197}
\label{Fig-inside-outsdie-NMR}
\vspace{-15mm}
\caption{Fig.~(a), Grey scale plot of resistivity versus magnetic
field and carrier concentration.  Fig.~(b), Model of the
ferromagnetic phase transition caused by different subband LL
spin alignment. Arrows denote spin orientations of LLs close to
$\nu=6$ expressed by dot-dashed line.  Fig.~(c), (d), (e) show
NMR signals close to various positions dotted in Fig.~(a).
Notice that NMR is only detected inside ring structure (Fig.~d),
and absent on its left (Fig.~c) and right side (Fig.~e).
NMR at a particular $B$ was measured for several different
values of $V_g$. For a comparison, theoretical NMR values,
assuming $\gamma=7.29$ MHz/T for $^{75}$As,
are shown by dashed vertical lines.}
\end{figure}

Fig.~2a shows the phase diagram of longitudinal resistivity,
$\rho_{xx}$, as a function of magnetic field, $B$,
and carrier density, $n$, at 70 mK.
Its most salient feature is the ring-like structure
centered around the $\nu=6$ line. It was previously speculated that
the ring represents a ferromagnetic state realized by preferential
alignment of electron spins of different subbands,
when separation between LLs of different subbands is
comparable with the exchange interaction energy
$E_{ex}$ \cite{Zhang2005PRL}. The schematic diagram for this model
is illustrated in Fig.~2b. On either side of the ring
the system is in a paramagnetic state with exactly the
same quantized Hall conductance, 6$e^2/h$,
because the topmost two occupied LLs are two spin split
states of the same subband. Recently the ring-like structures
have been numerically simulated by Ferreira
\it{et al.}$\;$\rm \cite{Ferre2006XXX}.

To probe the aforementioned regions with a means other
than transport, resistively detected NMR was
performed in the proximity of the ring structure.
No NMR was seen on either side of the
ring structure, as shown in Fig.~2c and 2e. In contrast,
inside the ring structure pronounced NMR signals were observed.
An example is shown at $B$=4.55 T for four different values of $V_g$.
The relative change of $R_{xx}$ is typically about 1$\%$ at
resonance at an estimated RF power about 0.1 mW at the sample.
The resonant frequencies at different
$V_g$ agree well with the theoretical values,
indicated by the vertical dashed line in Fig.~2d.
Upon resonance, $R_{xx}$ in all NMR lines shows a sharp decrease
followed by a much slower relaxation
process back to its original value, which is characterized by
the nuclear spin-lattice relaxation time constant, $T_1$,
as will be discussed.

\begin{figure}[t]
\begin{center}
\epsfig{file=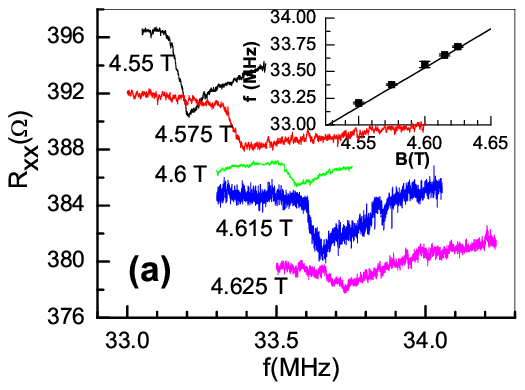, width=0.77\linewidth, angle=0}
\epsfig{file=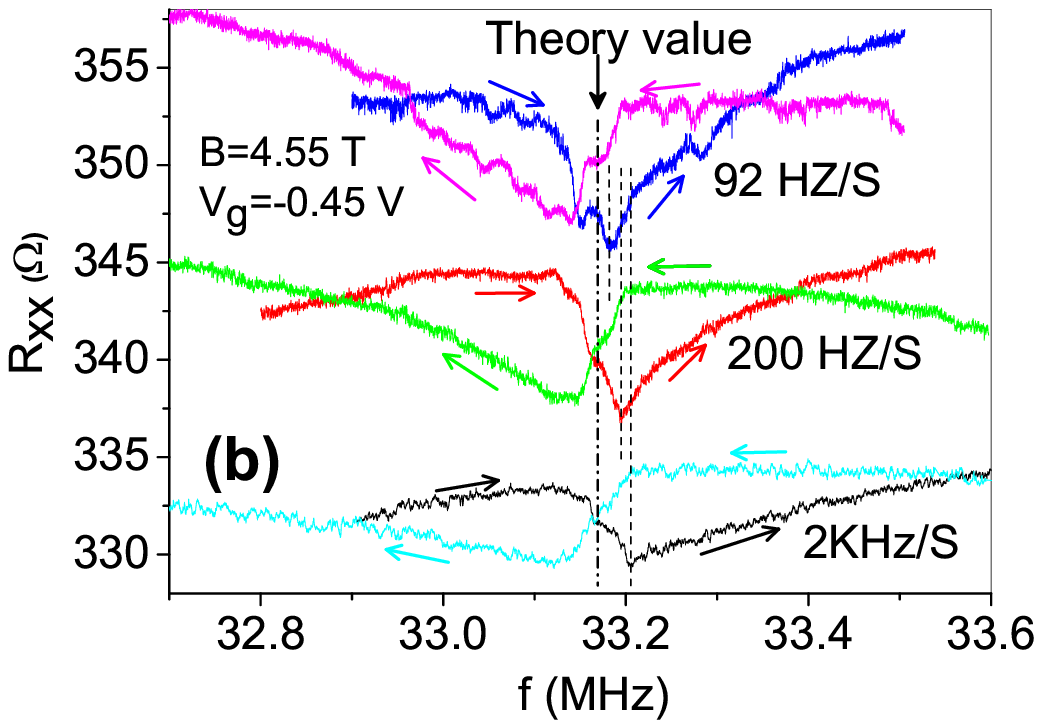, width=0.82\linewidth, angle=0}
\end{center}
\vspace{-5mm}
\caption{Fig.~(a), NMR resonances at different magnetic fields
inside the ring structure. Inset shows the resonance frequency as
a function of magnetic field. The straight line has a slope of
7.29 MHz/T for $^{75}$As resonance. Fig.~(b), detected NMR signals
at $B=4.55$ T and $V_g=-0.45$ V, at three frequency scanning speeds,
and in opposite sweeping direction (indicated by arrows).
The theoretical resonance frequency at 33.1695 MHz is
indicated by dot-dashed vertical line.  The curves in both figures
are shifted along y-axis for clarity.}
\label{Fig-NMR-B}
\end{figure}

To further confirm the NMR signal, NMR resonance lines were recorded
at different magnetic fields inside the ring structure,
as shown in Fig.~\ref{Fig-NMR-B}a.  The resonance
shows the expected blue shift with increasing $B$.
The frequencies of the minima in the NMR lines, plotted in the inset,
exhibit a linear relationship with $B$, and are slightly above
the theoretical values denoted by the line.
Nevertheless, Fig.~\ref{Fig-NMR-B}b
reveals that the NMR line form is strongly dependent on frequency sweeping
direction and speed.  The slower the frequency sweeps,
the closer the minima frequency approaches the expected theoretical value.
Resistively detected NMR towards the ring center is hampered
by the exponentially vanishing $R_{xx}$ value,
similar to the earlier study of the Skyrmion state around
$\nu=1$ \cite{Gervais2005PRL}.

We believe the mechanism of resistively detected NMR here is
similar to that described in the literature. For the 2D electron
system in GaAs, the contact hyperfine interaction between the
nuclear spin $\mathitb {I}$ and the electron spin
$\mathitb {S}$ can be expressed as
$\mathitb {A I \cdot S}=\frac{A}{2}(I^+S^-+I^-S^+)+AI_zS_z$,
where $A$ is the hyperfine coupling constant \cite{Dobers1988PRL}.
Due to the term $\frac{A}{2}(I^+ S^- + I^- S^+)$,
a nuclear spin flops, $\downarrow \Rightarrow \uparrow$,
when an electron spin flips, $\uparrow \Rightarrow \downarrow$.
Nuclear spins that have once flopped hardly relax back because of
their longer relaxation time $T_1$, which is on the order of minutes,
relative to that of the electrons.  Hence up-spin nuclei pile up
to develop a strong nuclear polarization $I_z$, which is parallel
to an external magnetic field $B$. It will reduce the effective
electron spin flip energy,
\begin{equation}
E_z=g^* \mu_B BS_z + A\langle I_z \rangle S_z,\label{Zeeman}
\end{equation}
as $g^*<0$ \cite{Desrat2002PRL}. When the NMR resonance condition
is matched, the nuclear spins are depolarized and
the electron Zeeman energy increases consequently.
Since $R_{xx}$ is dependent on the energy
gap of the energy spectrum, $\Delta E$, and thermally activated according
to $R_{xx} \propto exp(-\Delta E/2kT)$, the NMR is manifested by a
drop in $R_{xx}$, as shown by all the
NMR lines in Fig.~1-3.

The observed prominent (absent) NMR signal inside (outside) the ring
structure in Fig.~2c-e, is well correlated with the spin polarization
of the picture depicted in Fig.~2b \cite{Zhang2005PRL}. Inside the
ring along the $\nu=6$ line, the topmost two occupied LLs from different
subbands have the same spin orientation and hence $S_z = 1$ in Eq.~(1).
Accordingly, NMR signal was observed in Fig.~2d.  Whereas outside the
ring along the $\nu=6$ line, it is a normal paramagnetic state where spin
split LLs of different subbands are completely filled. Hence $S_z=0$
in Eq.~(1), and NMR cannot be expected.
 
\begin{figure}[t]
\begin{center}
\epsfig{file=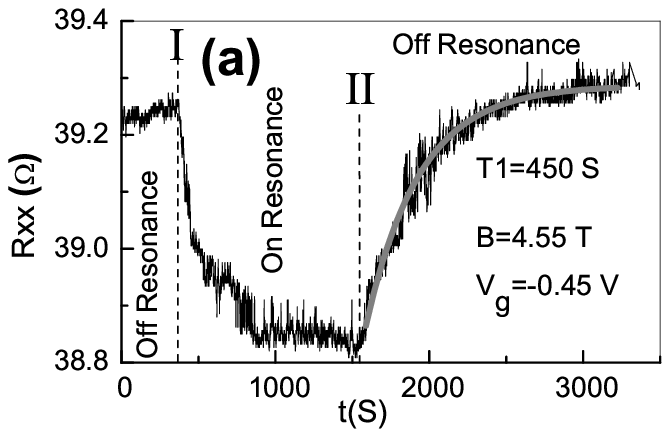, width=0.82\linewidth, angle=0}
\epsfig{file=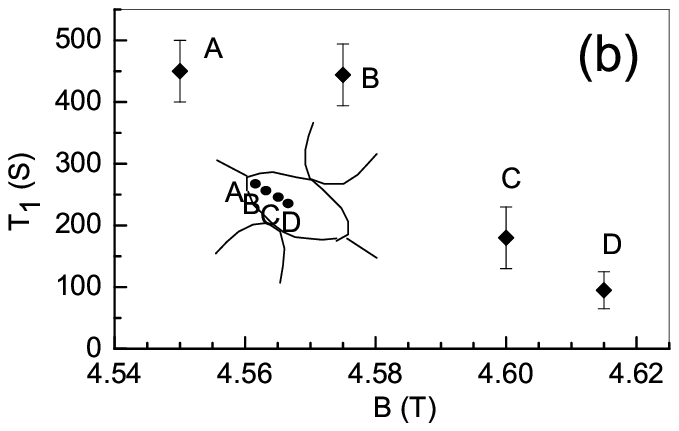, width=0.86\linewidth, angle=0}
\end{center}
\vspace{-5mm}
\caption{Fig.~(a) shows the principle of measuring $T_1$ by recording time
evolution of $R_{xx}$ irradiated by RF, initially off resonance, on
resonance (position I), and finally off resonance (position II). The
on-resonance frequency was set to 33.1695 MHz.  $T_1$ is determined by
an exponential fit to the experimental data shown by
the thick gray curve.  Fig.~(b), acquired $T_1$ for points shown in its
inset exhibits a declining tendency towards the center of the ring structure.
Inset is a zoom-in view of the ring structure.  Dots A, B, C, and D show a
path leading towards the ring center,
A(4.55 T, -0.45 V), B(4.575 T, -0.459 V), C(4.6 T, -0.468 V),
D(4.615 T, -0.475 V).}
\label{Fig-Relax}
\end{figure}

To gain more insight into the nature of the state inside the ring, we studied
the coupling between the nuclei and electrons by measuring the nuclear
spin-lattice relaxation time, $T_1$, at various positions inside the ring
structure. Fig.~4a shows the procedure for determining $T_1$ by imitating the
nuclear polarization regaining process after complete depolarization. Initially
the sample is irradiated by an off-resonance RF.
At I, RF was tuned into resonance, and
$R_{xx}$ decays exponentially into a steady state,
which means that nuclei have been completely depolarized. Then at II, the
frequency was switched back to off-resonance, nuclei gradually restore their
polarization owing to the interaction with the electron spin bath, and
consequently $R_{xx}$ slowly relaxes back to its original value. This
relaxing process can be well fitted by an exponential function of the
form $R_{xx}=\alpha + \beta e^{(-t/T_1^{\prime})}$ with a time constant
of $T_1^{\prime}$, which is approximately equal to the nuclear
spin-lattice relaxation time $T_1$ if $g\mu_B B_N \ll 2k_B T$ and the
RF power is small \cite{Gervais2005PRL}.

The obtained $T_1$ is plotted in Fig.~4b as a function of $B$ for
different locations inside the ring structure marked by dots in the
inset picture. These positions, with their successively diminishing
resistance, represent a path progressively leading to the center
of the ring structure. Each data point is an average of several
repeated measurements. $T_1$ rapidly drops from 450
to 95 s towards the ring center, indicating
a more efficient nuclear spin relaxation.
Following $R_{xx}$, we found that $T_1$ reduces as electrons
become more localized. This correlation of $T_1$ with
localization is \it{opposite}$\;$\rm to the usual Korringa
relation in metals, $1/T_1 \propto D(E) \propto \rho_{xx}$,
$D(E)$ is density of states.
However, the data share a remarkable
resemblance with that observed by
Gervais \it{et al.}$\;$\rm \cite{Gervais2005PRL}. In that study,
the fast relaxation rate was attributed to the localized
Skyrmion crystal relaxing the nuclear spin via a noncollinear spin wave
Goldstone mode \cite{Cote1997PRL}. Stimulated by this assertion,
we intend to suggest that the electrons inside the ring form a
many-body state with collective spin excitations. Therefore,
the localized electrons can provide an efficient relaxation
path for nuclear spins.
The theoretically predicted spin-density-wave instabilities
near the paramagnetic-ferromagnetic transition \cite{Giuli1985PRB}
can potentially be a source of these quantum fluctuations.

We also found that measurements up to 400 mK with the same
frequency sweeping speed reveal negligible change in the
line shape of the NMR signal. Assuming a Boltzmann distribution,
at $T=400$ mK and $B=4.55$ T, the percentage of
polarized $^{75}As$, $\eta$, is only $0.2\%$.  The corresponding
resonance amplitude $\Delta R_{xx}/R_{xx}=\Delta E_z/2kT$
is about $0.2\%$, where $\Delta E_z=g^*\mu_BB_N$, and the
nuclear field generated by the polarized $^{75}As$
nuclei $B_N = -1.84 \langle I^{^{75}As} \rangle=-1.84 \cdot (3/2 \cdot \eta)$
in Tesla \cite{Dobers1988PRL}. This is nearly one order of magnitude
lower than the experimental value of $1\%$.  Thus the observed
nuclear spin polarization cannot be built thermally and
it has to be gained by some dynamic means.

Indeed, it was found that electrical current can efficiently
induce electron spin-flip and nuclear spin-flop processes in the
$\nu=2/3$ fractional quantum Hall spin-polarized
state \cite{Kronmuller1998PRL, Kraus2002PRL}.
Since a ferromagnetic state can spontaneously separate
into domains, as the applied current forces
electrons to scatter between adjacent domains with different spin
but almost degenerate energy, the nuclei in the neighborhood
can become polarized.
Adapting this picture of dynamic polarization, we can imagine
that strong nuclear polarization observed at high temperatures
in our experiment is consistent with the fact that
the ring state consists of spontaneously polarized electron spins.

In conclusion, resistively detected NMR was detected only inside
the ring-like structure in the phase diagram of a two-subband
electron system. The associated nuclear spin-lattice relaxation
time decays rapidly towards the center of the ring and the NMR
signal persists to temperatures as high as 400 mK. These findings
are consistent with the notion that there is a ferromagnetic state
inside the ring structure facilitated by preferential alignment
of electron spins from different subband in an external magnetic field.
 
The authors would like to thank G. Clark for helpful discussions,
and B. Alavi for technical assistance. This work is supported
by NSF under grant DMR-0404445.

\end{document}